\documentclass{an}
\usepackage{times}
\usepackage{graphicx,aabib99}
\begin{document}


\title{Velocity dispersion in N-body simulations of CDM models}
 
\author{Alvaro Dom\'\i nguez
\institute{Max-Planck-Institut f\"ur Metallforschung, Heisenbergstr.
  3, D--70569 Stuttgart, Germany}}

\date{29 April 2003}

\abstract{This work reports on a study of the spatially coarse-grained
  velocity dispersion in cosmological N-body simulations (OCDM and
  $\Lambda$CDM models) as a function of time (redshifts $z=0$--$4$)
  and of the coarsening length ($0.6$--$20\ h^{-1}$Mpc). The main
  result is the discovery of a polytropic relationship ${\cal I}_1
  \propto \varrho^{2-\eta}$ between the velocity-dispersion kinetic
  energy density of the coarsening cells, ${\cal I}_1$, and their mass
  density, $\varrho$.
  The exponent $\eta$, dependent on time and coarsening scale, is a
  compact measure of the deviations from the naive virial prediction
  $\eta_{\rm virial}=0$. This relationship supports the ``polytropic
  assumption'' which has been employed in theoretical models for the
  growth of cosmological structure by gravitational instability.
   }

\correspondence{alvaro@fluids.mpi-stuttgart.mpg.de}

\maketitle

\section{Introduction}

Some recent works
\cite{BDGP97,BuDo98,AdBu99,DHMP99,BDP99,MTM99,Domi00,MoTa01,Domi02,TSMM02}
have explored a formulation of hydrodynamic kind for the formation of
cosmological structures by gravitational instability.
It intends to describe the dynamical evolution of the few most
relevant fields (typically the coarse-grained mass density and
velocity fields) in terms of a set of autonomous equations for those
fields, much in the same way as the hydrodynamic equations for usual
fluids.
The widely used {\it dust model} \cite{Peeb80,SaCo95} belongs to this
class, but it has shortcomings, most noticeably the emergence of
singularities, beyond which its application is invalid. An interesting
result of the systematic study of the hydrodynamic formulation are the
corrections to the dust model. They arise from the nonlinear coupling
of the evolution to the structure below the coarsening length, and
turn out to be relevant in the nonlinear regime ($\leftrightarrow$ the
regime of large density fluctuations about homogeneity). The corrected
equations are generalizations of the {\it adhesion model}
\cite{KoSh88,GSS89,KPSM92,MSW94,SSMP95}, which is able to reproduce
successfully the gross features of the structure evolved by
gravitational instability in different cosmological scenarios. The
hydrodynamic formulation requires the corrections to dust to be
expressed as functions of the coarse-grained mass density and velocity
fields, so as to get a closed set of equations for these fields. This
is, however, a major theoretical problem, since one cannot follow the
standard procedure to close the hydrodynamic hierarchy by invoking
``local thermal equilibrium'' \cite{Huan87,Bale91}: the ideas and the
formalism of thermodynamics cannot be applied straightforwardly to a
system dominated by its own gravity, this being in fact still an open
question (see, e.g., the concise review by Hut \cite*{Hut97}).

In this work I report on the empirical search for closure
relationships with the help of N-body simulations of large-scale
structure formation. I consider in particular the velocity dispersion
of the particles contained in any coarsening cell, which shows up in
the equation for momentum conservation (in the hydrodynamic parlance,
the kinetic contribution to the pressure and to the viscous stresses).
The simulation results for the trace of the velocity dispersion (the
internal kinetic energy density of the coarsening cells) can be fitted
quite well by a ``polytropic relationship'', borrowing the terminology
from thermodynamics: ${\cal I}_1 \propto \varrho^{2-\eta}$.
This result supports the polytropic approximation used in the
theoretical derivation of adhesion-like models \cite{BDP99}. The
polytropic relationship improves with time, and so I conclude that it
must be a consequence of the evolution by gravitational instability.
For small coarsening lengths or large times, the exponent $\eta$ is
close to, but significantly different from the virial prediction
following from the simple model of isolated, relaxed, structureless
halos.

The degree of anisotropy of the velocity dispersion (the departure
from a spherically symmetric distribution of the dispersion) has also
been studied, but in this case the scatter of the data is large and
the results do not provide a significant conclusion. The anisotropy
tends to decrease with increasing mass density (till a few percents at
the largest probed densities), being however always larger than that
associated to a Maxwellian distribution with the same density. This
decrease at least speaks in favor of the approximation of isotropic
velocity dispersion also employed in deriving adhesion-like models.

This work is organized as follows: in Sec.~\ref{secmet} I detail the
method employed in the analysis of the simulations. Sec.~\ref{secres}
presents the results, which are discussed with the help of theoretical
arguments in Sec.~\ref{secdis}. Finally, a brief mathematical
discussion of some topics required in the main text are collected in
the Appendix.

\begin{figure}
  \resizebox{\hsize}{!}{\includegraphics{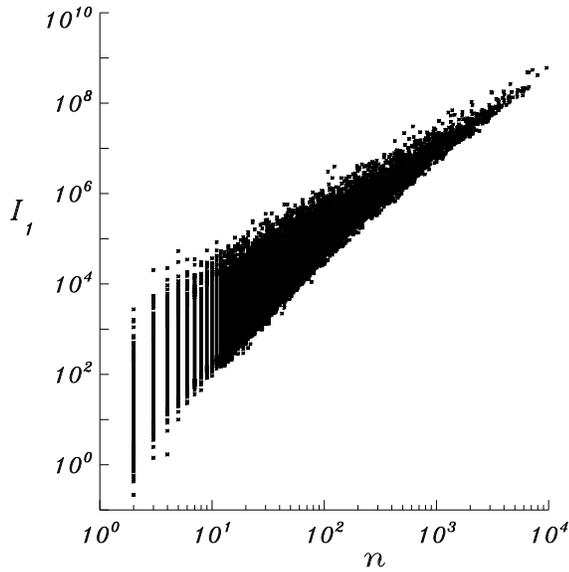}}
  \caption{
    Typical log-log scatter plot of the trace of the velocity
    dispersion tensor (in units of km$^2$ s$^{-2}$ (Mpc$/h)^{-3}$,
    setting the particle mass $m=1$) vs the number of particles. The
    data correspond to the $\Lambda$CDM model at redshift $z=0$ with a
    smoothing length $L=4.55\ h^{-1}$Mpc.}
  \label{figraw}
\end{figure}

\section{Method} \label{secmet}

The analyzed CDM simulations were performed by {\it The Hydra
  Consortium} \cite{CTP95} using the $AP^3M$ algorithm. They consist
of a cubic box (periodic boundary conditions) of side length $100\ 
h^{-1}$Mpc at the present epoch, containing $N=86^3$ particles. Two
different cosmological models have been considered, OCDM ($\Omega_m =
0.3$, $\Omega_\Lambda = 0.0$, $h=0.81$, $\sigma_8 = 1.06$, $\Gamma =
0.25$, $m=1.63 \times 10^{11}\ M_{\sun}$) and $\Lambda$CDM ($\Omega_m
= 0.3$, $\Omega_\Lambda = 0.7$, $h=0.96$, $\sigma_8 = 1.22$, $\Gamma =
0.25$, $m=1.37 \times 10^{11}\ M_{\sun}$), at three times,
corresponding to redshifts $z=0$, $1.4$, $3.6$.

The hydrodynamic formulation can be derived by means of a coarsening
procedure (Dom\'\i nguez 2000, 2002). Given a comoving smoothing scale $L$,
the coarse-grained mass density, velocity and velocity dispersion
fields are defined respectively as follows:
\begin{displaymath}
  \varrho ({\bf x}, t) = {m \over [a(t) L]^3} \sum_{\alpha=1}^{N} \, 
  W \left({{\bf x}-{\bf x}_\alpha \over L} \right) , 
\end{displaymath}
\begin{equation}
  \label{fields}
  \varrho {\bf u} ({\bf x}, t) = {m \over [a(t) L]^3} 
  \sum_{\alpha=1}^{N} \, {\bf u}_\alpha(t) \, 
  W \left({{\bf x}-{\bf x}_\alpha \over L} \right) ,
\end{equation}
\begin{eqnarray*}
  \Pi ({\bf x}, t) = {m \over [a(t) L]^3} \sum_{\alpha=1}^{N} & & 
  [{\bf u}_{\alpha}(t) - {\bf u}({\bf x}, t)] \otimes \\
  & & [{\bf u}_{\alpha}(t) - {\bf u} ({\bf x}, t)] \, 
  W \left({{\bf x}-{\bf x}_\alpha \over L} \right) .
\end{eqnarray*}
($\Pi$ is a second-rank tensor, $\otimes$ denoting a dyadic product).
Here, $a(t)$ denotes the expansion factor, $m$ is the mass of a
particle, ${\bf x}_\alpha$ and ${\bf u}_\alpha$ represent the comoving
position and peculiar velocity, respectively, of the $\alpha$-th
particle, and $W(\cdot)$ is a (normalized) smoothing window. Exact
dynamical equations can be derived for $\varrho$ and ${\bf u}$,
expressing mass and momentum conservation; the latter equation
features the velocity dispersion in the form of a term $\nabla \cdot
\Pi$. The purpose of the present study is to check if there exists any
approximate relationship between $\Pi$ and the field $\varrho$, so
that an autonomous set of equations can be written for the fields
$\varrho$ and ${\bf u}$.

Starting from the coordinates $\{ {\bf x}_\alpha, {\bf u}_\alpha
\}_{\alpha=1 \dots N}$ provided by the simulation, the definitions
(\ref{fields}) were implemented with a cubic top-hat window,
\begin{equation}
  \label{window}
  W({\bf z}) = \theta (1 - 2 |z_1|) \; \theta (1 - 2 |z_2|) \; 
  \theta (1 - 2 |z_3|) ,
\end{equation}
where $\theta(\cdot)$ is the step function. 
In total 13 different values of $L$ were explored, spanning the range
from $0.6 \ h^{-1}$Mpc up to $20 \ h^{-1}$Mpc and equally separated in
a logarithmic scale. For each value of $L$, 
at least $2 \times 10^4$ randomly centered, non empty
coarsening cells were probed.

\begin{figure*}
  \resizebox{\hsize}{!}{\includegraphics{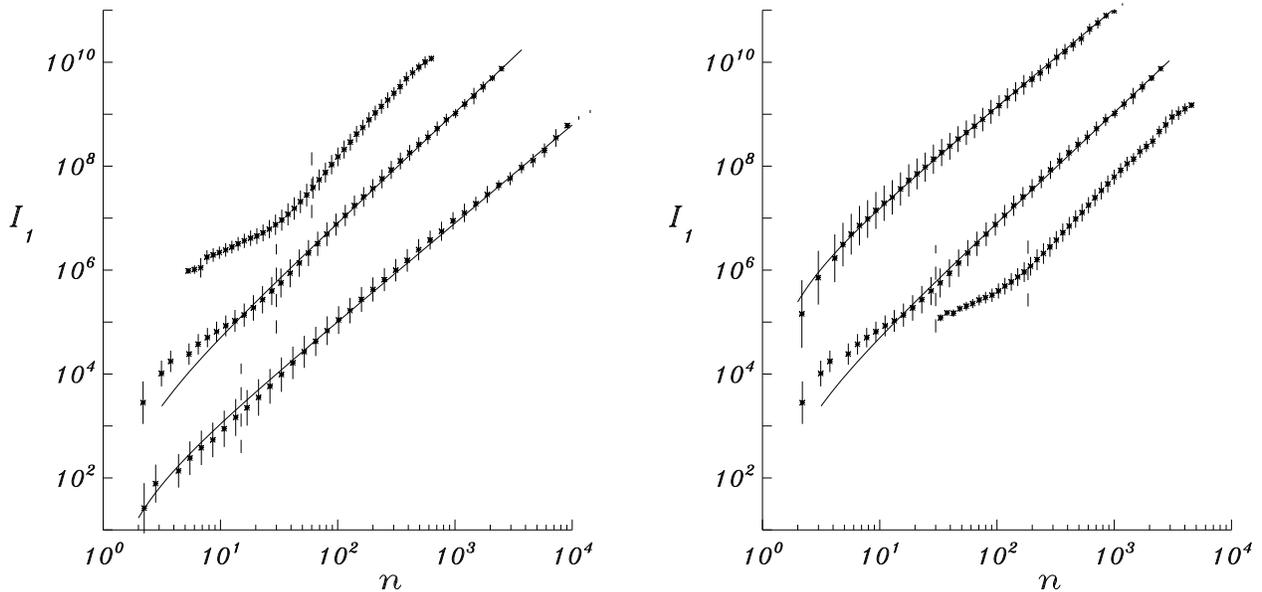}}
  \caption{
    Binned log-log plot of the trace of the velocity dispersion tensor
    (same units as Fig. \ref{figraw}) vs the number of particles for
    the $\Lambda$CDM model. The error bars correspond to the estimated
    $1$-$\sigma$ variance. On the left, for three redshifts (from top
    to bottom, $z=3.6$, $1.4$, $0$) at a fixed smoothing length
    $L=4.55\ h^{-1}$Mpc. On the right, for three smoothing scales
    (from top to bottom, $L=1.41$, $4.55$, $8.33\ h^{-1}$Mpc) at a
    fixed redshift $z=1.4$.  The solid line corresponds to the
    fit~(\ref{poly}), and the dashed line marks $n_c$,
    Eq.~(\ref{ncrit}).}
  \label{figbin}
\end{figure*}

The velocity dispersion tensor $\Pi_{ij} ({\bf x})$ was studied as a
function of the density $\varrho ({\bf x})$, or equivalently, of the
number of particles contained in the cell at ${\bf x}$, namely $n
({\bf x}) = [(a L)^3 / m] \varrho ({\bf x})$. For the purposes of this
work, it sufficed to consider the three eigenvalues $\lambda_i$ of the
tensor $\Pi$, or better, its three principal scalar invariants:
\begin{displaymath}
  {\cal I}_1 = \textrm{tr} \; \Pi = \lambda_1 + \lambda_2 + \lambda_3 ,
\end{displaymath}
\begin{displaymath}
  {\cal I}_2 = {1 \over 2} \left[ (\textrm{tr} \; \Pi)^2 - 
    \textrm{tr} \; (\Pi : \Pi) \right] = \lambda_1 \lambda_2 + 
  \lambda_2 \lambda_3 + \lambda_3 \lambda_1 ,
\end{displaymath}
\begin{equation}
  \label{invariants}
  {\cal I}_3 = \textrm{det} \; \Pi = \lambda_1 \lambda_2 \lambda_3 .
\end{equation}
The quantity $(1/2) {\cal I}_1$ is the peculiar kinetic energy per
unit volume due to the motion of the particles relative to the cell
center of mass.
The other two invariants can be related to the degree of anisotropy of
the tensor $\Pi$. More precisely, the dimensionless coefficients
\begin{equation}
  \label{aniso}
  \alpha = 1 - {3 \, {\cal I}_2 \over {\cal I}_1^2} , \qquad 
  \beta = 1 - {27 \, {\cal I}_3 \over {\cal I}_1^3} ,
\end{equation}
quantify the departure from the isotropic case,
$\lambda_1=\lambda_2=\lambda_3=p$. I show in the Appendix that $0 \leq
\alpha, \beta \leq 1$, and they vanish if and only if $\Pi$ is
isotropic. Moreover, it follows from its definitions that $\beta
\approx 3 \alpha$ for small anisotropy.

To check that this smoothing algorithm was right, it was applied to an
ideal gas simulation.
The results for the dependence of the kinetic pressure $p=(1/3) {\cal
  I}_1$ and the anisotropy parameters $\alpha$, $\beta$ on $n$ agree
with the predicted relationships (see the Appendix).

In the next Section I also discuss how the results depend on the
smoothing details (e.g., on the choice of the window (\ref{window})).
I anticipate that the conclusions to be extracted are robust.

\section{Results} \label{secres}

The results concerning the three invariants as a function of the
particle number are qualitatively the same in the whole range of
smoothing scales and times explored, and for the two cosmological
models OCDM and $\Lambda$CDM. Fig.~\ref{figraw} is a typical example
of a ${\cal I}_i$ vs $n$ scatter plot. The data suggest a polytropic
relationship between the scalar invariants and $n$, more and more
acurate for larger $n$. Hence, it appears that the increasingly larger
scatter for smaller particle numbers is mainly due to the discrete
character of the variable $n$. To eliminate this noise, the log-log
plots of the raw data were binned into 40 subintervals on the
$n$-axis: this effectively means averaging the scalar invariants for
different fixed values of $n$ and provides also an estimate of the
variance. This scatter is in fact the main source of error in the
parameters $\eta$ and $\kappa$ of the fit~(\ref{poly}).
Fig.~\ref{figbin} collects a set of representative cases after
implementation of this procedure, where the advocated polytropic
dependence is evident. To be sure that this averaging method does not
introduce artificial features, the analysis was repeated by varying
the number of bins, from 10 up to the limiting case in which each
possible value of the discrete variable $n$ is treated as a bin. The
conclusions are robust and the results do not depend on the amount of
binning. (Changes of the number of bins within this range induce
variations in the best-fit parameters $\eta$ and $\kappa$ which are
well within the error region in Fig.~\ref{figparam}). The choice of 40
bins is a good compromise that efficiently discards the noise but
still preserves the relevant features of the ${\cal I}_i$-$n$
relationship.

\begin{figure}
  \resizebox{\hsize}{!}{\includegraphics{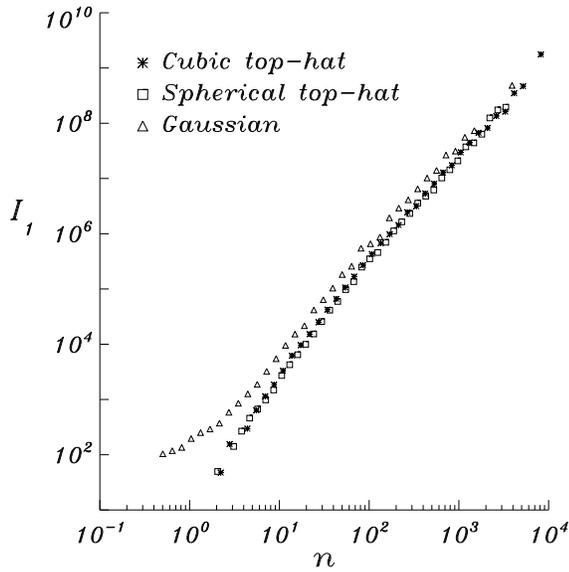}}
  \caption{
    Log-log plot of the trace of the velocity dispersion tensor vs the
    number of particles, computed with three different smoothing
    windows. The data correpond to the $\Lambda$CDM model, $z=0$,
    $L=3.45\ h^{-1}$Mpc.}
  \label{figwindows}
\end{figure}

Another check probed the influence of the choice of smoothing window.
The differences found between different windows could be explained as
a consequence of the discrete nature of $n$. Fig.~\ref{figwindows}
shows the typical result after using three different smoothing
windows: a cubic top-hat, Eq.~(\ref{window}), a spherical top-hat,
$W({\bf z}) = \theta (1 - (4 \pi/3)^{1/3}|{\bf z}|)$, and a Gaussian,
$W ({\bf z}) = \exp (-\pi |{\bf z}|^2)$. The windows are normalized to
unity and give the same weight (=1) to the origin: the comparison is
then straightforward. The cubic and spherical top-hat windows yield
indistinguishable results, demonstrating that window anisotropy is not
relevant (this could also be expected, see the last paragraph in this
Sec.). The Gaussian smoothing also gives similar results, but now a
slight difference in the small-$n$ end is observed, because the
discreteness of $n$ implies (i) that $\Pi=0$ if $n=1$ for the top-hat
windows and (ii) that for small $n$, $\Pi$ is dominated by far
contributions from the Gaussian tails.
However, as discussed later, the discreteness constraint $\Pi=0$ if
$n=1$ for a top-hat window can be taken into account in a simple
manner and indeed the polytropic fit~(\ref{poly}) holds very well even
for $n \sim 1$. Hence, the cubic top-hat window was selected, because
it is also most easily implemented.

A final check, carried out only for the particular time $z=0$, was to
restrict the coarsening procedure to a subvolume of the simulation box
(1/64 of the total volume), so as to find out to what extent the
results are exclusive of the simulation volume of $100\ h^{-1}$Mpc
side length: as expected, the quality of the fits are slightly worse
because of the reduced number of particles, but the conclusions remain
the same.

\begin{figure*}
  \resizebox{\hsize}{!}{\includegraphics{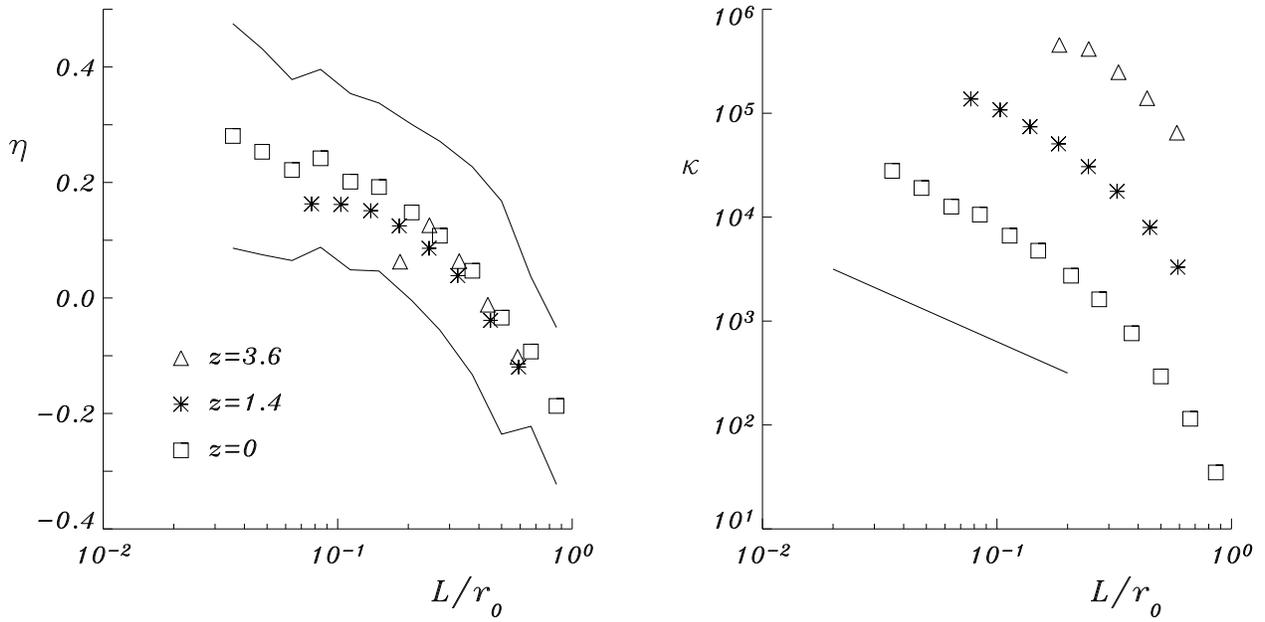}}
  \caption{
    Plot of the best-fit parameters in Eq.~(\ref{poly}) vs the
    smoothing length for the $\Lambda$CDM model at the three times
    studied; $\kappa$ in units of km$^2$ s$^{-2}$. The scale of
    nonlinearity is $r_0$, Eq.~(\ref{sigma1}).  The two lines in the
    $\eta$-plot delimit the estimated $1$-$\sigma$ error region at
    time $z=0$ (the errors are of the same size at the other times and
    they have been omitted for clarity). The error bars for $\kappa$
    are about the same size as the plot symbols. The virial prediction
    reads $\eta=0$ and $\kappa \propto 1/L$ (solid line in the
    $\kappa$-plot).}
  \label{figparam}
\end{figure*}

As exemplified by Fig.~\ref{figbin}, one finds a well defined
relationship between ${\cal I}_1$ and $n$, in which three different
behaviors can be identified: a polytropic dependence ${\cal I}_1
\propto n^{2-\eta}$ for $n$ larger than a certain value $n_c$, a
bending upwards for $1<n<n_c$, and finally a bending downwards for $n$
close to 1 which is due to the constraint $\Pi=0$ if $n=1$. The
intermediate behavior is absent at late times/small coarsening
lengths, when it is masked by the latter constraint. The value of
$n_c$ increases with the redshift and the smoothing length. A rough
estimate by eye of the function $n_c (L,z)$ is given by
\begin{equation}
  \label{ncrit}
  n_c=F(z) \bar{n}_L , \qquad \bar{n}_L:=(86 \, L /100)^3 .
\end{equation}
$\bar{n}_L$ is the average particle number in a cell of comoving side
length $L$ (in units of $h^{-1}$Mpc), and $F(z)$ is a mild function of
the redshift, $F=1$ for $z=3.6$ and decreases slowly for smaller $z$
($F=0.25$ for $z=0$). This fit for $n_c (L,z)$ is conservative in the
sense of slightly overestimating $n_c$ as $L$ decreases, however it is
precise enough: Fig.~\ref{figbin} shows that the polytropic fit is in
fact still well followed by the data for a certain range below the
chosen $n_c$. The best-fit parameters $\eta$ and $\kappa$ in
Eq.~(\ref{poly}) are quite insensitive to the precise value of $n_c$.
The intermediate range $1<n<n_c$ corresponds to underdense cells
($n_c$ is associated to a density contrast $\delta_c = F(z)-1$); the
deviations from the polytropic fit may then be a remnant of the
artificial discreteness, most evident in the lattice structure of the
initial conditions.
Nonetheless, this failure of the polytropic relationship is likely
unimportant from a dynamical point of view, being restricted to ever
more rarefied regions ($\delta_c \approx -0.75$ at $z=0$).

The binned data in the range $n>n_c$ were fitted to the polytropic
relationship
\begin{equation}
  \label{poly}
  {\cal I}_1 = {m \over (a L)^3} \, \kappa \, (n-1)^{2-\eta} . 
\end{equation}
The factor $n-1$ enforces the
discreteness constraint ${\cal I}_1=0$ for $n=1$. This improves the
fit on the small-$n$ region in the cases that it extends down to $n
\sim 1$ and it is also suggested by the simulated ideal gas: the ideal
gas pressure-density relation is still obeyed with the replacement $n
\rightarrow n-1$ when the smoothing length is so small that most of
the coarsening cells contain just a few particles. The parameters
$\eta$ and $\kappa$ were determined by the least-squares method. The
fit was carried out only when the fit range in $n$ spanned at least a
decade (in which case the range in ${\cal I}_1$ always extends over at
least two decades); the polytropic fit is still consistent when this
condition is not met, but the large errors deprive it of significance.
In the best cases, the fit range in $n$ included three decades. The
time and smoothing-length dependences of the parameters $\eta$ and
$\kappa$ for the $\Lambda$CDM model are plotted in
Fig.~\ref{figparam}. The results for the OCDM model are very close.

\begin{figure}
  \resizebox{\hsize}{!}{\includegraphics{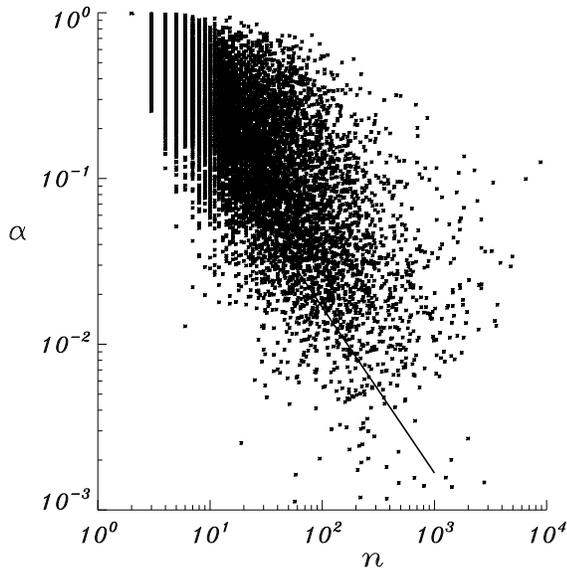}}
  \caption{
    Typical log-log scatter plot of the anisotropy parameter $\alpha$
    vs the particle number. The data correspond to the $\Lambda$CDM
    model, $z=0$, $L=4.55\ h^{-1}$Mpc. The straight line represents the
    average anisotropy for a Maxwellian distribution.}
  \label{figaniso}
\end{figure}

The other two invariants, ${\cal I}_2$ and ${\cal I}_3$, are also
fitted by a polytropic dependence. In fact, it turns out that the
anisotropy parameters $\alpha$ and $\beta$ {\em computed with the
  fitting functions} are extremely small and $n$-independent. The
reason is that the binned (=averaged) velocity dispersion, being only
a function of the {\em scalar} $n$, must be isotropic. The anisotropy
parameters must be calculated for the raw data, before binning:
Fig.~\ref{figaniso} shows a typical $\alpha$ vs $n$ scatter plot. The
plots of the parameter $\beta$ look similar (in fact, the relation
$\beta \approx 3 \alpha$ is a very good approximation already for
$\alpha \sim 0.1$: then, one should rather study, e.g., $\beta - 3
\alpha$ to gain non-redundant information).  Apart from the large
scatter for small particle number due to the discreteness of $n$,
there is also an important dispersion even for large $n$, because the
anisotropy must be determined by something else than only a scalar.
Moreover, unlike the ``extensive'' scalar invariants (= sums of
positive contributions from many particles), the anisotropy
parameters, being essentially the difference between eigenvalues,
Eq.~(\ref{eigenaniso}), are more sensitive to the discreteness of $n$.
Thus, the data quality just allows to confirm that $\alpha$ and
$\beta$ tend to decay with increasing $n$ but to remain somewhat
larger than the average ideal gas anisotropy, $\alpha_\mathrm{ideal}
\approx 5/(3 n)$ (see the Appendix). No significant dependence (if
any) with time and smoothing length could be detected.

\section{Discussion and conclusion} \label{secdis}

The simulations show a clear evidence for a polytropic relationship
(\ref{poly}) between the density and the scalar invariants of the
velocity dispersion tensor. The fits improve with decreasing redshift
or smoothing scale; hence one is lead to view them as a general
consequence of the evolution by gravitational instability. In
agreement with this interpretation, this relationship occurs for both
the $\Lambda$CDM and the OCDM models. Thus, it seems that the
existence of the polytropic dependence itself is independent of the
background cosmological model, which would perhaps only affect the
values of the fitting parameters slightly.

The theoretical explanation of the precise values of the fitting
parameters is not evident. Fig.~\ref{figparam} shows that the function
$\eta(L, z)$ can be approximately written in fact as a function of the
single variable $L/r_0(z)$, where $r_0 (z)$ is the scale of
nonlinearity, defined by the condition that the variance of the
density contrast is unity:
\begin{equation}
  \label{sigma1}
  \frac{\langle (n - \bar{n}_L)^2 \rangle}{\bar{n}_L^2} = 1, \qquad 
  \textrm{for $L=r_0$.}
\end{equation}
The $\Lambda$CDM simulation provides the values $r_0=3.2, 7.7, 16.7\ 
h^{-1}$Mpc, respectively for the three considered redshifts. This
property agrees nicely with the evolution by gravitational instability
in a hierarchical scenario. The points for $\kappa(L, z)$ can be made
to collapse on a single curve too if $\kappa$ is also suitably
rescaled by a $z$-dependent factor; but this must be viewed as pure
phenomenology, since no theoretical explanation for the values of this
rescaling factor could be given.

A theoretical argumentation by Buchert and Dom{\'\i}nguez
\cite*{BuDo98} provides a polytropic relationship with a fixed
$2-\eta=5/3$: this is equivalent to the adiabatic evolution of an
ideal gas and was justified for early times and under restrictive
initial conditions. However, Fig.~\ref{figparam} shows that, precisely
in the opposite limit of nonlinear scales/large times ($L<r_0$),
$\eta$ is close to this ``adiabatic value'', which even seems to be an
asymptote. But the errors are too large and the probed range of
nonlinear lengths too narrow to draw a firm conclusion.

This flattening of $\eta (L)$ suggests another possible explanation of
the polytropic relationship. It seems sensible to hypothesise that,
for smoothing lengths $L$ well in the nonlinear regime, the kinetic
energy should be fixed by the virial theorem: if the coarsening cells
can be idealized as virialized, structureless halos, then $(a L)^3
{\cal I}_1$ should be proportional to the potential gravitational
energy of the coarsening cell, which could in turn be estimated as
$\sim G (m n)^2/a L$. This implies immediately a polytropic
dependence~(\ref{poly}) with $\eta_{\rm virial}=0$ and $\kappa_{\rm
  virial} \propto (a L)^{-1}$. This idealized model was employed to
simulate scatter plots as Fig.~\ref{figraw}: assume exact isotropy
($\alpha=\beta=0$) and the same distribution of particle numbers $n$
as obtained in the simulations, and compute ${\cal I}_1$ from the
condition that the differences ${\bf u}_\alpha-{\bf u}$ in the
definition~(\ref{fields}) of $\Pi$ are independent, Gaussian
distributed random variables with zero mean and a variance given by
the virial theorem. In this way, an estimated $\sigma < 2 \cdot
10^{-3}$ for $\eta_{\rm virial}$ was gained.  Fig.~\ref{figparam}
shows that the best-fit values of $\eta$ and $\kappa$ are close to the
virial predictions for small $L$; but the deviations of $\eta$ are
well above the estimated fluctuations in $\eta_{\rm virial}$.

The reason for this discrepancy must lie on the assumptions involved
in the above reasoning: (i) the potential energy should be dominated
by the contribution from the structure on the scale $L$ and (ii) each
coarsening cell should be approximately isolated and in a relaxed,
stationary state, so that the virial theorem holds. The long-range
nature of gravity could perhaps justify assumption (i) even in a
hierarchical scenario, when the matter distribution is far from smooth
inside the coarsening cells, provided it is not too diluted on the
scale $L$ either. But it cannot be excluded that the corrections due
to the substructure contribute significantly to the discrepancy.
Assumption (ii) can be easily violated: the coarsening cells may have
a significant interaction with neighboring cells, or be part of a
larger virialized halo or contain a bunch of streaming particles far
from any (quasi-)stationary state. As evidence, consider Fig.~3 in
\cite{KnMu99}: it is equivalent to my ${\cal I}_1$-$n$ plots but the
points correspond to groups determined by a friends-of-friends
algorithm, rather than by coarsening boxes of fixed side length. What
Knebe and M\"uller identify as unvirialized groups clearly tend to
yield a positive $\eta$, in agreement with the trend observed in
Fig.~\ref{figparam}. Therefore, a very interesting result is that the
violations to the ``virialized halo'' conditions (i-ii) do not destroy
the polytropic relationship itself predicted by the virial theorem,
but only change the values of the parameters.

In conclusion, N-body simulations have provided evidence for a
polytropic relationship between the coarse-grained mass density and
the peculiar kinetic energy. This relation can be characterized by an
exponent $\eta$ which measures the (scale and time dependent)
departures from the ``virialized halo'' prediction. It was found that,
for smoothing lengths well in the nonlinear regime, $\eta$ is
significantly larger than zero. There still remains the task of
theoretically explaining this polytropic dependence. Future work in
this direction will address scale-invariant cosmological models: they
provide testbed cases which can be easily controlled in simulations
and easily analyzed theoretically.

\appendix
\section{The anisotropy parameters $\alpha$ and $\beta$}

In this Appendix I derive some properties of the anisotropy parameters
defined in Eqs.~(\ref{aniso}). Since the tensor $\Pi$ is
positive-definite, then $\lambda_i \geq 0$. This implies the bounds $0
\leq \alpha, \beta \leq 1$.  In fact, ${\cal I}_i \geq 0$ yields
immediately that $\alpha$, $\beta \leq 1$. On the other hand, for any
triplet of non-negative numbers, the following inequalities hold
\cite{AbSt65}:
\begin{displaymath}
  \sum_{i=1}^3 \lambda_i^2 \geq \lambda_1 \lambda_2 + 
  \lambda_2 \lambda_3 + \lambda_3 \lambda_1 , 
\end{displaymath}
\begin{equation}
  \frac{1}{3} \sum_{i=1}^3 \lambda_i \geq  \left( \lambda_1 \lambda_2 
    \lambda_3 \right)^{1/3} ,
\end{equation}
and the equality is satisfied if and only if all three numbers are
equal. Combining these inequalities with the definitions
(\ref{invariants}-\ref{aniso}), it is found that $\alpha$ and $\beta$
must be non-negative, and they are zero if and only if the tensor
$\Pi$ is isotropic.

Let us write $\lambda_i = \frac{1}{3} {\cal I}_1 + \delta \lambda_i$,
so that the coefficients $\delta \lambda_i$ represent the deviation
from isotropy and satisfy $\sum_i \delta \lambda_i = 0$. Then:
\begin{equation}
  \label{eigenaniso}
  \alpha = - \frac{3}{{\cal I}_1^2} \sum_{i<j} 
  (\delta \lambda_i) (\delta \lambda_j) ,
\end{equation}
\begin{displaymath}
  \beta = 3 \alpha - \frac{27}{{\cal I}_1^3} 
  (\delta \lambda_1) (\delta \lambda_2) (\delta \lambda_3) ,
\end{displaymath}
and in the limit of small anisotropy, $|\delta \lambda_i| \ll {\cal
  I}_1$, the equality $\beta \approx 3 \alpha$ holds.

The simulations of the ideal gas provide $\langle
\alpha_\mathrm{ideal} \rangle_n \approx 5/(3 n)$ with a large scatter,
for the reason discussed towards the end of Sec.~\ref{secres}
($\langle \cdots \rangle_n$ refers to the binning procedure detailed
in that Sec.). Nevertheless, this result is reliable because the $1/n$
dependence can be easily explained: $\Pi$ for a coarsening cell
containing $n$ particles is the sum of $n$ independent random
variables. Hence, for not too small $n$, the average $\langle \Pi
\rangle_n$ will be isotropic and extensive, with the scaling ${\cal
  I}_1 \sim n$, while fluctuations around isotropy will scale like
$|\delta \lambda| \sim \sqrt{n}$. Expression (\ref{eigenaniso}) then
yields that $\alpha \sim 1/n$.

\acknowledgements 

The author is grateful to P.\ A.\ Thomas for providing the simulations
data, and to C.\ Beisbart, T.\ Buchert, R.\ Dom\'\i nguez--Tenreiro,
M.\ Kerscher, P.\ A.\ Thomas and R.\ Trasarti--Battistoni for their
useful comments on the manuscript. The major part of this work was
done at the Laboratorio de Astrof\'\i sica Espacial y F\'\i sica
Fundamental, Madrid, and the Ludwig-Maximilians-Universit\"at,
M\"unchen.

\end{document}